\documentclass[twocolumn, twocolappendix]{aastex631}

\usepackage{amsmath}
\usepackage{amssymb}
\usepackage{bm}
\usepackage{xcolor}

\received{November 8, 2024}
\accepted{January 11, 2025}

\submitjournal{ApJ}

\newcommand{\brackets}[1]{\left[#1\right]}

\graphicspath{{./}{figs/}}
\begin{document}

\title{The Guitar Filament's Magnetic Field Revealed by Starlight Polarization}

\author[0000-0002-6401-778X]{Jack T. Dinsmore}
\affiliation{Department of Physics, Stanford University, Stanford CA 94305, USA}
\affiliation{Kavli Institute for Particle Astrophysics and Cosmology, Stanford University, Stanford CA 94305, USA}
\author[0000-0001-6711-3286]{Roger W. Romani}
\affiliation{Department of Physics, Stanford University, Stanford CA 94305, USA}
\affiliation{Kavli Institute for Particle Astrophysics and Cosmology, Stanford University, Stanford CA 94305, USA}
\author[0000-0002-2567-2132]{Nikos Mandarakas}
\affiliation{Institute of Astrophysics, Foundation for Research and Technology-Hellas, 71110 Heraklion, Greece}
\affiliation{Department of Physics, University of Crete, 70013 Heraklion, Greece}
\author[0000-0003-0611-5784]{Dmitry Blinov}
\affiliation{Institute of Astrophysics, Foundation for Research and Technology-Hellas, 71110 Heraklion, Greece}
\affiliation{Department of Physics, University of Crete, 70013 Heraklion, Greece}
\author[0000-0001-9200-4006]{Ioannis Liodakis}
\affiliation{Institute of Astrophysics, Foundation for Research and Technology-Hellas, 71110 Heraklion, Greece}

\begin{abstract}
  The Guitar nebula surrounding PSR B2224+65 boasts a pulsar X-ray filament likely aligned with the local magnetic field. We present new RoboPol stellar polarization data distributed along the line-of-sight to the pulsar. The polarizing effect of intervening magnetized dust allows us to extract a model for the dust-weighted magnetic field. We detect a magnetic field angle consistent with the filament if the pulsar is located in the more distant zone of its parallax-estimated distance range.
\end{abstract}

\section{Introduction} \label{sec:intro}
\begin{figure}
  \centering
  \includegraphics[width=\linewidth]{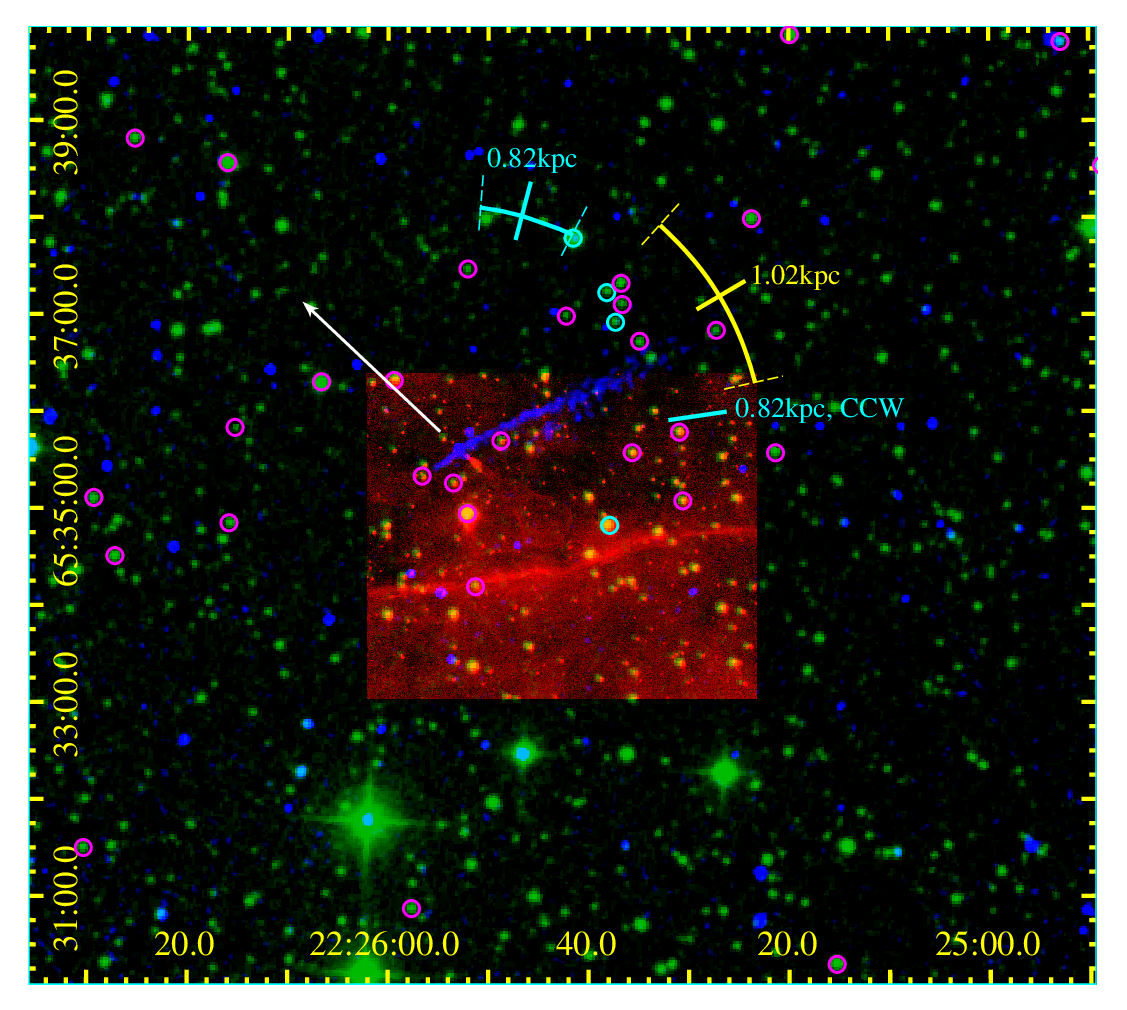}
  \caption{
  The Guitar field in equatorial coordinates showing the H$\alpha$ bow shock and a trailing H$\alpha$ front (red), optical stars (green) and \textit{Chandra} soft X-ray image including the filament (blue). The pulsar proper motion is marked with the white arrow. RoboPol-observed stars are marked (magenta) with a few X-ray-bright stars excluded (cyan); others lie outside the frame. The cyan and yellow arcs mark the inferred ISM field direction (and 1$\sigma$ errors) for $d=0.82$\,kpc and $d=1.02$\,kpc CW solutions, respectively. The isolated cyan bar marks the center of the 0.82\,kpc CCW solution. 
  }
  \label{fig:guitar}
\end{figure}

The Guitar H$\alpha$ nebula trailing PSR B2224+65 \citep{cordes1993guitar} sports a long, narrow X-ray ``filament'' \citep{2007A&A...467.1209H} at large angle to the pulsar's well-measured transverse velocity [$760\,(d/0.83\,\mathrm{kpc})$\,km s$^{-1}$; Fig.~\ref{fig:guitar}]. With this obliquity, the filament cannot be a PWN trail. Four additional pulsar X-ray filaments have since been discovered, along with a few candidates. They generally feature large pulsar velocity, low Galactic latitude, and hard X-ray spectral indices \citep[photon index $\Gamma \approx 1.6$,][]{dinsmore2024catalog}.

A likely explanation is that the filament consists of synchrotron-radiating, ultrarelativistic electrons and positrons accelerated near the pulsar to Lorentz factors of $\sim 10^{8}$. When the large pulsar velocity and/or large local ISM density compress the bow shock stand-off, the leptons may escape and stream along the external ISM magnetic field lines. The physical properties of the filament pulsars support this picture \citep{bandiera2008on, dinsmore2024catalog}, as do numerical simulations of the pulsar environment \citep{olmi2019full}.
This model requires an ambient magnetic field aligned with the filament position angle. \cite{churazov2024pulsar} show promising evidence for such alignment in the filament candidate G0.13$-$0.11, using X-ray polarization data from the Imaging X-ray Polarimetry Explorer (\textit{IXPE}). However, the object's lack of a confirmed pulsar and its location near the crowded Galactic center complicate the interpretation.

For other objects, we can test the model by measuring the ambient field orientation. The correlation between stellar polarization and the local magnetic field direction enables this. Asymmetric dust grains show polarization-dependent absorption and align with the ISM magnetic field. Thus, unpolarized starlight gains polarization parallel to the plane-of-sky component of the magnetic field at the few percent level \citep{andersson2015interstellar, davis1951polarization}.

Here we present observations of the plane-of-sky direction of the magnetic field local to Guitar using stellar polarization vectors as measured by RoboPol \citep{ramaprakash2019robopol} at the Skinakas Observatory. The analysis separates the polarizing effect of dust clumps along the line-of-sight (LoS) \S \ref{sec:methods}. While the results are not yet definitive, the measurements are consistent with the expected field orientation at the pulsar's distance \S \ref{sec:results}. We note how additional measurements can refine this test and discuss briefly the prospects for tests of other filaments \S\ref{sec:conclusion}.

\section{Methods}
\label{sec:methods}

\begin{figure}
  \centering
  \includegraphics[width=\linewidth]{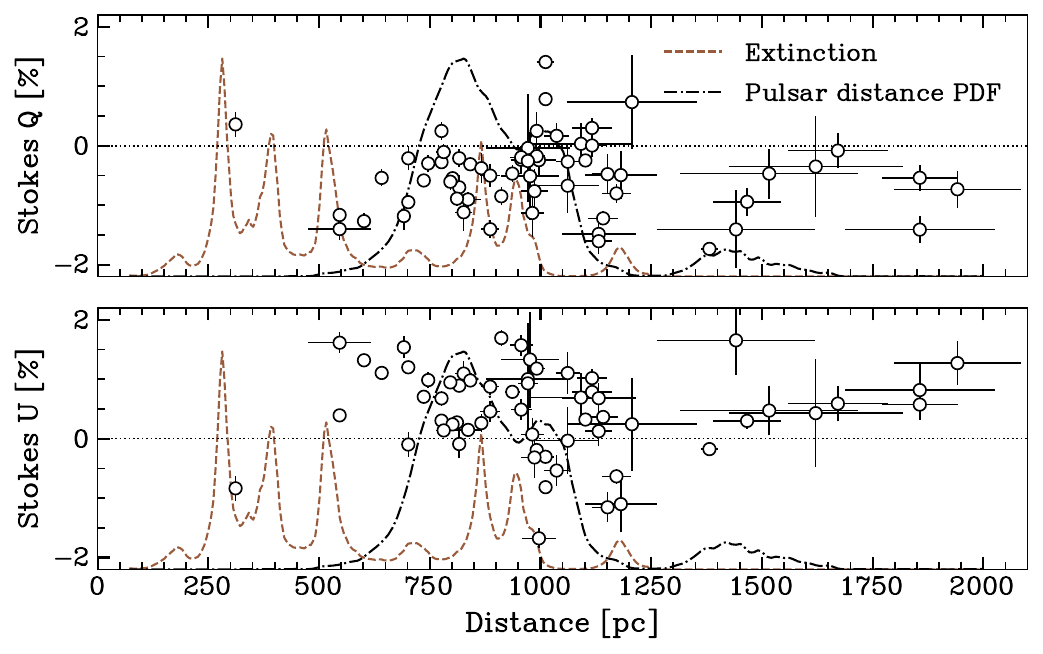}
  \caption{Stellar polarizations measured as a function of distance. Dust density as measured by extinction (brown) and the pulsar distance PDF (black) are also shown.}
  \label{fig:stars}
\end{figure}

Here we measure polarization angle $\psi$ counterclockwise from North, defining the Stokes coefficients $Q = \Pi\cos 2\psi$ and $U = \Pi\sin 2\psi$ for polarization degree $\Pi$.

\subsection{Data}
\label{sec:data}

We conducted RoboPol polarization measurements of stars in the \textit{Gaia} DR3 catalog \citep{gata2021data} within 10$'$ of the center of the Guitar filament with $G_{RP}<17$. We required parallax distances $d=$0.3--2\,kpc (with $\sigma_d < 0.2$\,kpc) bracketing the Guitar's VLBI parallax distance of $\sim 0.82$\,kpc \citep{deller2019microarcsecond}. A few targets were excluded due to crowding. Of the 65 stars selected for observation, four stars are coincident with \textit{Chandra} X-ray point sources; since these may exhibit intrinsic polarization, they were dropped from the analysis. The remaining 61 stars (Fig.~\ref{fig:guitar}) were processed with the standard RoboPol pipeline \citep{2014MNRAS.442.1706K}, delivering estimates of Stokes $Q$ and $U$ and uncertainties (Fig.~\ref{fig:stars}).

The three-dimensional map of Galactic dust \citep{edenhofer2023parsec} provides important constraints on the dust-induced polarization. This map reveals eight differential stellar extinction peaks in  $dE/d\ell$ out to $\ell=1.25$ kpc (see Fig.~\ref{fig:stars}), with $\ell$ precision $\approx 50$ pc from the width of the peaks. Two peaks lie near Guitar's estimated distance; we chose stars bracketing each peak to constrain its polarizing effect. The map predicts $dE/d\ell$ along each star's LoS, used in the analysis as described in the next section.

\subsection{Statistical Analysis}
In order to extract the most information from our relatively small data set, we write a full likelihood for the detected Stokes coefficients and constrain the magnetic field structure using Bayesian techniques. We assume that each star's detected polarization is drawn from a Gaussian distribution whose mean  and variance are related to the magnetic field structure as modeled below.

On average, a star's polarization is proportional to its extinction \citep{fosalba2002statistical, panopoulou2019demonstration, 2023A&A...670A.164P}, implying expected Stokes coefficients of
\begin{align}
  Q(\ell) &= \eta \int_0^\ell d\ell'\, \frac{dE}{d\ell} \cos [2\phi(\ell')]\\ 
  U(\ell) &= \eta \int_0^\ell d\ell'\, \frac{dE}{d\ell} \sin [2\phi(\ell')]
\end{align}
where $\eta$ is the polarizing efficiency and $\phi$ is the plane-of-sky magnetic field angle. By integrating over all distances from 0 to $\ell$, we account for the foreground effects of dust between the star and the telescope.

For fields lying in the plane of sky, our $\eta$ parameter would be approximately the maximum $9\%$ found in \cite{fosalba2002statistical}. Fields with a LoS component will have smaller $\eta$. Our sample size is too small to provide good constraints on $\eta$ in individual dust clumps, which could be caused by varying magnetic inclination. We therefore assume $\eta$ is constant and fit for the global mean value. In effect, our method is otherwise similar to that described in \cite{2023A&A...670A.164P}, except our complex LoS dust distribution leads us to use a continuous $B$ model constrained by dust maps of $dE/d\ell$.

The $Q,\,U$ data show large, random fluctuations, suggesting that the Stokes coefficients possess some intrinsic variance $\sigma_B(\ell)^2$. This has been attributed to turbulence randomizing the local magnetic field orientation \citep{chandrasekhar1953magnetic,pelgrims2024first}. Since the fluctuations arise from the magnetic field, their variance should be proportional to extinction just as the expected $Q$ and $U$ from the constant component are. Thus,
\begin{equation}
  \sigma_B(\ell)^2  = \int_0^\ell d\ell'\, \frac{dE}{d\ell} \delta^2
\end{equation}
where $\delta$ characterizes the fluctuation amplitude. We apply this uncertainty to both Stokes coefficients, so that the standard deviation of our Gaussian likelihood is the quadrature sum of $\sigma_B$ and the measurement uncertainty. To be precise, the log-likelihood of each stellar polarization measurement is
\begin{equation}
\begin{aligned}
  \ln L_s(\ell_s) = -&\frac{1}{2}\bigg[\frac{(Q(\ell_s) - Q_s)^2}{\sigma_{Q,s}^2 + \sigma_B(\ell_s)^2} + \frac{(U(\ell_s) - U_s)^2}{\sigma_{U,s}^2 + \sigma_B(\ell_s)^2}\\
  +& \ln (\sigma_{Q,s}^2 + \sigma_B(\ell_s)^2) + \ln (\sigma_{U,s}^2 + \sigma_B(\ell_s)^2)\bigg]
  \label{eqn:chisq}
\end{aligned}
\end{equation}
where the subscript $s$ runs over the star list. The distribution of $-2\ln L_s$ calculated from the data under the best-fit model closely follows the $\chi^2$ distribution with 1 degree of freedom, validating this uncertainty model.

Since each star's distance is uncertain, our final fit likelihood $L$ is obtained by marginalizing the likelihood of Eq.~\ref{eqn:chisq} over $\ell_s$:
\begin{equation}
  \ln L = \sum_{s=1}^{N_\mathrm{stars}} \ln \int d\ell_s\, P(\ell_s) L_s(\ell_s).
  \label{eqn:likelihood}
\end{equation}
The distance probability
\begin{equation}
  P(\ell_s) \propto \ell^2_s\exp\brackets{-\frac{(\pi - 1/\ell_s)^2}{2\sigma_\pi^2}}
  \label{eqn:distance-pdf}
\end{equation}
for \textit{Gaia}-measured parallax $\pi$ accounts for the increased prior for the star to be located at large $\ell$ (the Lutz-Kelker bias).

\begin{figure}
  \centering
  \includegraphics[width=\linewidth]{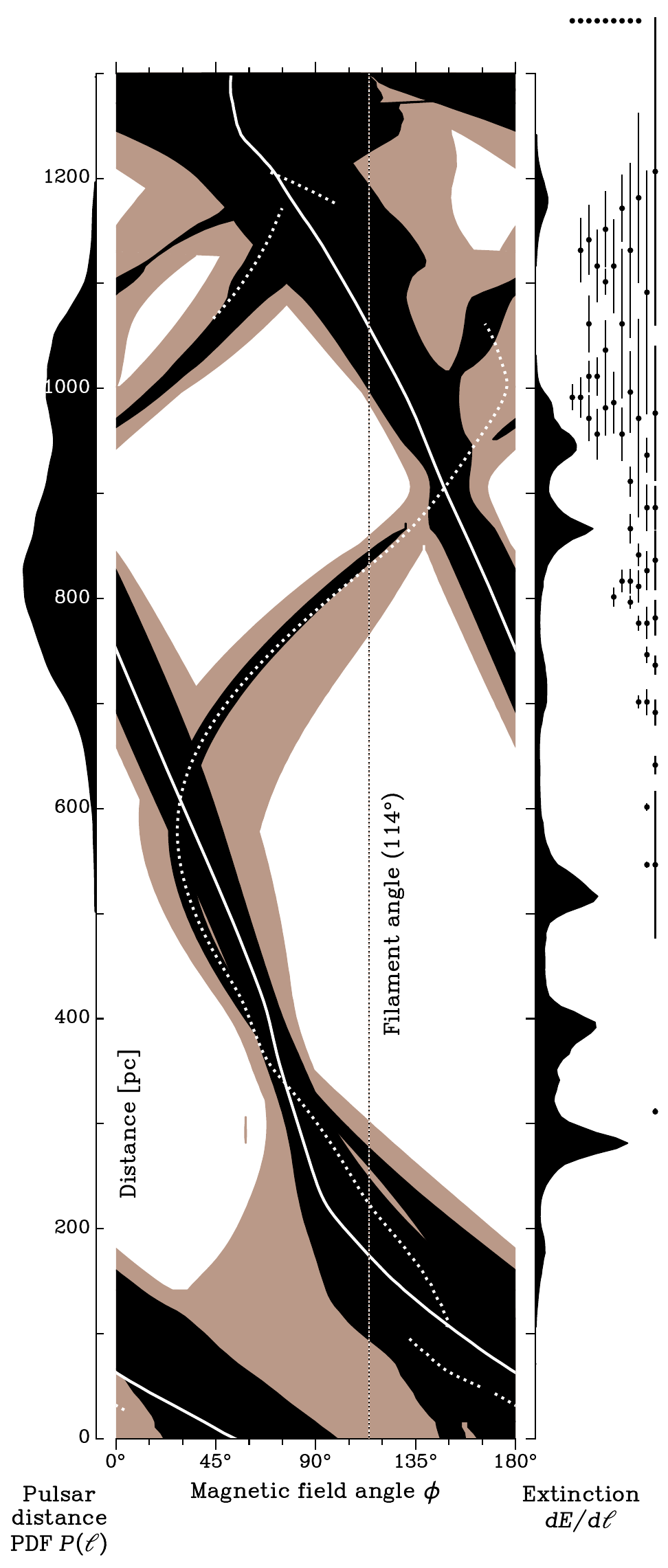}
  \caption{\emph{Center}: 68\% (black) and 95\% (brown) posterior PDF contours for the plane-of-sky magnetic field angle as a function by distance. White lines show the locations of the two main posterior modes (solid CW; dotted CCW). \emph{Left:} The pulsar distance PDF. \emph{Right}: LoS differential extinction and the surveyed star distances and uncertainties.}
  \label{fig:results}
\end{figure}

We parameterize the magnetic field model as a spline running from 0 to 1.25 kpc, with the spline values at five equally spaced nodes $\phi_1,\dots,\phi_5$.  In principle, the field might take arbitrary excursions between the data-constrained nodes. However we restrict to the simplest, smoothly connecting models. Of these, only two have significant support from the data: one with the principle wrap clockwise (CW, solid line in Fig.~\ref{fig:results}), the second counterclockwise (CCW, dotted line). We fit for magnetic field angles in these two modes using the Python \texttt{emcee} implementation of the affine invariant Markov chain Monte Carlo (MCMC) ensemble sampler \citep{foreman2013emcee}. Wide priors are assumed for $\delta$ and $\eta$. We then merge the two fits by selecting a subset from their combined samples, which we evolve using the Metropolis-Hastings algorithm until convergence. The final samples are shown in Fig.~\ref{fig:results}.

\section{Results and Discussion}
\label{sec:results}

The Guitar LoS includes two dense dust layers: a ``0.4\, kpc'' ($0.3-0.5$\,kpc) layer foreground to the filament and a ``0.9\,kpc'' ($0.85-0.95$\,kpc) layer near the filament. Our data precisely constrain the magnetic field angle in these two regions (as shown by the tight contours around $0.4$ kpc and $0.9$ kpc in Fig.~\ref{fig:results}) to angles $\sim 90^\circ$ apart. In the poorly constrained region between the two layers, we see the CW and CCW modes mentioned above. The weak extinction layer at $\sim 0.7$ kpc slightly prefers the CW solution (solid line; 69\% of the posterior lies in this mode), though it does not exclude CCW (dotted line).

As present, our conclusions are limited by the pulsar parallax uncertainty. The most likely 0.82\,kpc distance lies just inside the 0.9\,kpc cluster. Here, the dust-inferred $\phi={166^\circ}^{+12}_{-13}$ in the dominant CW mode is inconsistent with the filament. However, the parallax measurement shows a secondary mode at 1.02 kpc, where the $\phi={123^\circ}^{+19}_{-17}$ CW mode measurement is consistent with the filament to $<1\sigma$ (see Fig.~\ref{fig:guitar}). The secondary CCW winding is consistent with the filament at the 2$\sigma$ level near 0.82 kpc but inconsistent at 1.02 kpc. If we marginalize over the uncertain pulsar distance, we find $\phi_\mathrm{Guitar}={167^\circ}_{-43}^{+34}$. Our best-fit $\eta$ parameter suggests a magnetic field inclined approximately $45^\circ$ from the LOS on average.

The \textit{Gaia}-based \texttt{StarHorse} stellar distance catalog \citep{anders2022photo} provides updated distance estimates for most of the stars in our dataset, based on the stars' physical properties. Since these distances include non-parallax constraints, their PDFs should be more complicated than the simple form we use (Eq.~\ref{eqn:distance-pdf}), so we do not use them for our main results. However, when \texttt{StarHorse} distances replace the raw \textit{Gaia} parallax distances, the $\phi$ constraint in both parallax modes moves closer to the filament by $\lesssim 0.5 \sigma$; the 1.02\,kpc mode moves to 114$^\circ$ (lying on the filament), while the 0.82\,kpc mode falls to 162$^\circ$.

\section{Conclusion}
\label{sec:conclusion}

Our current picture of pulsar filaments requires an aligned field to magnetically duct the synchrotron emitting leptons, but measuring the local field orientation is a challenge. When possible, one uses the polarized emission of the filament itself to measure the plane of sky field. For the X-ray bright filament candidate G0.13$-$0.11, this was done with {\it IXPE}. Similar measurements may be possible for a few other bright filaments, but most are much too faint. In principle, radio polarization can provide information, and the radio filaments adjacent to G0.13$-$0.11 in the Galactic center support the X-ray orientation. 
However, this will not generally be a useful path as the confirmed X-ray filaments are not radio detected.
Measurements of the ambient ISM field, as pursued here for the Guitar nebula, are thus particularly attractive. 

Our tomographic dust-weighted measurement for the Guitar shows that there are multiple components to the intervening LoS field. We have modeled the large scale field structure, assuming smooth variations below the 100pc scale and a roughly constant magnetic field inclination angle, and we find that the orientation is consistent with that of the filament, especially if the actual distance lies at the upper end of the range allowed by current pulsar parallax measurements. More stellar polarization measurements would help, beating down the magnetic field angle uncertainty as $\sim \sigma_B / \sqrt{N}$ and allowing us to estimate inclination angle variations between dust clumps. Polarization precision is not particularly important, as long as it is subdominant to the large intrinsic dispersion $\sigma_B$. Thus, the future for filament studies likely lies with large area stellar polarization surveys rather than the targeted observations employed here.

When dust-weighted $\phi(\ell)$ are consistent over a large range bracketing the pulsar distance, interpretation is simple. When, as for the Guitar, the orientation varies strongly along the LoS, high precision pulsar distances are essential. For the Guitar, the present parallax measurements have a relatively poor $\chi^2$/DoF=3.6 with several outliers. Additional VLBI epochs can better separate the parallax and proper motion components and improve the distance determination, allowing better use of the local dust measurements.

Of the five confirmed filaments in \cite{dinsmore2024catalog}, three are covered by the \cite{edenhofer2023parsec} maps. After the Guitar, J2030+4415 looks to be interesting with dust layers bracketing the best guess pulsar distance. However this $\gamma$-ray only pulsar lacks a parallax, so if these layers show differing $\phi$, interpretation will be difficult. PSR J2055+2539 is very close at $d\sim 0.4$\,kpc, and only features one foreground dust layer with modest extinction. When large scale surveys \citep[e.g.\,PASIPHAE;][DragonFlyPol]{2018arXiv181005652T} become available, analysis of local polarization should be a useful probe of these and yet to be discovered pulsar filaments. 

\begin{acknowledgments}
The authors thank Adam Deller for providing valuable insight and data on the Guitar parallax measurement. This work was supported by Smithsonian Astrophysical Observatory grant 2777277 SAO\_GO.
\end{acknowledgments}

\vspace{5mm}
\software{\texttt{emcee} \citep{foreman2013emcee}, \texttt{dustmaps} \citep{green2018dustmaps}, \texttt{astropy} \citep{astropy2022astropy}, \texttt{Python}}

\bibliography{bib}{}
\bibliographystyle{aasjournal}

\end{document}